\documentclass[letterpaper,useAMS,usenatbib]{mn2e}
\usepackage{graphicx}
\usepackage{txfonts}

\def\xi{\hbox{$X_{\rm i}$}}
\catcode`\@=11
\def\gsim{\ifmmode{\mathrel{\mathpalette\@versim>}}
    \else{$\mathrel{\mathpalette\@versim>}$}\fi}
\def\lsim{\ifmmode{\mathrel{\mathpalette\@versim<}}
    \else{$\mathrel{\mathpalette\@versim<}$}\fi}
\def\@versim#1#2{\lower 2.9truept \vbox{\baselineskip 0pt \lineskip 
    0.5truept \ialign{$\m@th#1\hfil##\hfil$\crcr#2\crcr\sim\crcr}}}
\catcode`\@=12
\def\msun{\hbox{$M_\odot$}}

\def\t9{\hbox{$t_9$}}

\def\m*{\hbox{$M_*$}}

\def\ho{\hbox{$H_\circ$}}
\def\h50{\hbox{$\ho /50$}}

\def\y1{\hbox{${\rm yr}^{-1}$}}

\newcommand{\aap}{A\&A}
\newcommand{\mnras}{MNRAS}
\newcommand{\apj}{ApJ}
\newcommand{\aj}{AJ}

\title[Globular Cluster Formation]{The {\it Hubble Space Telescope} UV Legacy Survey of Galactic Globular Clusters. V. Constraints on Formation Scenarios.\thanks{Based on observations with  the
                               NASA/ESA {\it Hubble Space Telescope},
                               obtained at  the Space Telescope Science
                               Institute,  which is operated by AURA, Inc.,
                               under NASA contract NAS 5-26555.}
\author[A. Renzini et al.]
{A. Renzini,$^{1}$
F.\ D'Antona,$^{2}$
S.\ Cassisi,$^{3}$
I.\ R.\ King,$^{4}$
A.\ P.\ Milone,$^{5}$
P.\ Ventura,$^{2}$,
J.\ Anderson,$^{6}$
\newauthor
L.\ R.\ Bedin,$^{1}$
A.\ Bellini,$^{6}$
T.\ M.\ Brown,$^{6}$
G.\ Piotto,$^{7}$
R.\ P.\ van der Marel,$^{6}$
B.\ Barbuy,$^{8}$
\newauthor
E.\ Dalessandro,$^{9}$
S.\ Hidalgo,$^{10,11}$
A.\ F.\ Marino,$^{5}$
S.\ Ortolani,$^{7}$
M.\ Salaris,$^{12}$
and A.\ Sarajedini$^{13}$
\\
$^{1}${INAF-Osservatorio Astronomico di Padova, Vicolo
   dell'Osservatorio 5, I-35122 Padova, Italy}\\
$^{2}${INAF-Osservatorio Astronomico di Roma, Via Frascati
   33, I-00040 Monteporzio Catone, Roma, Italy}\\
$^{3}${INAF-Osservatorio Astronomico di Teramo, Via Mentore
   Maggini s.n.c., I-64100 Teramo, Italy}\\
$^{4}${Department of Astronomy, University of Washington,
  Box 351580, Seattle, WA 98195-1580}\\   
$^{5}${Research School of Astronomy and Astrophysics, The
  Australian National University, Cotter Road, Weston, ACT, 2611,
  Australia}\\   
$^{6}${Space Telescope Science Institute, 3700 San Martin
  Drive, Baltimore, MD 21218, USA}\\
$^{7}${Dipartimento di Fisica e Astronomia ``Galileo
  Galilei'', Universit\`a di Padova, Vicolo dell'Osservatorio 3, I-35122 Padova, Italy }\\
$^{8}${Universidade de S÷ao Paulo, IAG, Rua do Mat÷ao 1226, Cidade
  Universitaria, S÷ao Paulo 05508-900, Brazil}\\
$^{9}${Dipartimento di Fisica e Astronomia, Universit\`a di Bologna, Viale Berti Pichat 6/2, I-40127 Bologna, Italy}\\
$^{10}${Instituto de Astrofisica de Canarias, E-38200 La Laguna, Tenerife, Canary Islands, Spain}\\
$^{11}${Department of Astrophysics, University of La Laguna, E-38200 La Laguna, Tenerife, Canary Islands,
  Spain}\\
$^{12}${Astrophysics Research Institute, Liverpool John Moores University, Liverpool Science Park, IC2 Building, 146 Brownlow Hill,
  Liverpool L3 5RF, UK}\\
$^{13}${Department of Astronomy, University of Florida, 211
  Bryant Space Science Center, Gainesville, FL 32611, USA}\\
}}

\begin{document}

\date{Accepted September 29, 2015; Received July 8, 2015 in original form}

\maketitle
                                                            
\label{firstpage}

\date{}

\begin{abstract}
We build on the evidence provided by our Legacy Survey of Galactic globular clusters (GC) to submit to a crucial test four scenarios currently entertained for the formation of multiple stellar generations in GCs. The observational constraints on multiple generations to be fulfilled are manifold, including GC specificity, ubiquity, variety, predominance,
discreteness, supernova avoidance, $p$-capture processing, helium enrichment  and mass budget. We argue that scenarios appealing to supermassive stars, fast rotating massive stars and massive interactive binaries violate in an irreparable fashion two or more among such constraints. Also the scenario appealing to AGB stars as producers of the material for next generation stars encounters severe difficulties, specifically concerning the mass budget problem and the detailed chemical composition of second generation stars. We qualitatively explore ways possibly allowing one to {\it save} the AGB scenario, specifically appealing to a possible revision of the cross section of a critical reaction rate destroying sodium, or alternatively by a more extensive exploration of the vast parameter space controlling the evolutionary behavior of AGB stellar models. Still, we cannot ensure success for these efforts and totally new scenarios may have to be invented to understand 
how GCs formed in the early Universe.
\end{abstract}





\section{Introduction}
\label{sec:introduction}
Globular Clusters (GC) have always been among the most intensively studied stellar systems, and also among those whose images are so familiar to all astronomers.
Yet, we have never really  understood why it was so easy to produce such massive and extremely dense stellar aggregates, with over 100,000 stars per cubic parsec, and generate them in such a widespread fashion, no matter whether in a metal-poor or a metal-rich environment. We still have to understand how they formed, why they are so common around giant ellipticals  such as M~87, or spirals such as the MW Galaxy, or even in dwarfs such as Fornax or Sagittarius. Then, with the discovery of their multiple stellar populations, answering these questions became even harder than ever before. 

In this series of papers we report the results of the  {\it Hubble Space Telescope UV Legacy Survey of Galactic GCs}  dedicated to the observation of  48  GCs through the filters F275W, F336W, F438W of the Wide Field Camera 3 (WFC3)  on board {\it HST} (\citealt{piotto15}, hereafter Paper\,I ), which complements the existing F606W and F814W photometry from the Advanced Camera for Survey, or ACS (\citealt{sarajedini07}; \citealt{anderson08}) and is specifically designed to map multiple stellar populations in GCs. This set of filters, have proved to be most effective in disentangling the various sub-population, as these  bands sample different molecular absorptions, such as OH, NH, CH and CN, hence can distinguish stars with different degrees of $p$-capture processing (e.g., \citealt{milone12}).
Together with similar data already obtained in previous HST cycles,  our survey brings to 57 the number of GCs with such a homogeneous, multiband database which allows us to resolve such GCs in their multiple populations. As part of this project, the prototypical examples of M2 and NGC 2808 and that of  NGC 6352 have been already published (\citealt{milone15a,milone15b, nardiello15}, respectively paper II, III and IV). 

Photometric and spectroscopic evidence accumulated over the years has shown that secondary generations (2G) in GCs are made of materials enriched in helium, nitrogen and sodium
and depleted in carbon and oxygen, hence implying that they have been exposed to proton-capture reactions at high temperatures ($\sim 60-100$ million Kelvin), or more. Therefore, possible first generation (1G) sources   of the processed material (hereafter called the 1G donors) have included all kinds of stars  where $p$-capture materials are produced at such temperatures  and then likely ejected. Thus, the candidate producers of the raw 2G material  have been, in order of decreasing mass: i) supermassive stars
(SMS, $\sim 10^4\,\msun$, \citealt{denissenkov14}), ii) massive interacting binaries (MIB, e.g., $15+20\,\msun$, \citealt{demink09,bastian13}), iii) 
fast-rotating massive stars (FRMS, $\sim 25-120\,\msun$, \citealt{krause13}) and iv) asymptotic giant branch (AGB) and super-AGB stars (hereafter collectively, AGB, $\sim 3-10\,\msun$, e.g., \citealt{dercole10}).  Each of these candidate 1G donors is then part of its corresponding scenario depicting how the 1G-processed material might be incorporated into 2G stars. In the present paper we concentrate on exploiting the evidence presented in the earlier papers of this series to set constraints on these GC formation scenarios. In the last section we briefly mention other possible observational constraints.


\section{Binding Observational Constraints}

In Paper I we listed the main constraints on 2G formation that are imposed by the photometric and spectroscopic evidence. Here we recall them and add a few more.
\begin{itemize}
\item
{\bf GC Specificity.} The presence of 2G stars, with their chemical characteristics, is common within GCs, but stars with such characteristics are very rare in the Milky Way field. Their small number in the field is consistent with them having been generated within GCs and then lost by them through tidal interactions (e.g., \citealt{vesperini10,martell11}). This indicates that the (proto-)GC  environment may be indispensable for the production of  stars with such special composition.  In any event, every scenario aiming at accounting for GC multiple populations must at the same time account for the rarity of 2G-like stars in the field, i.e., it must be GC specific.  Young massive clusters (YMC, with mass up to $\sim 10^8\,\msun$) are occasionally forming  in the local Universe, but  do not appear to be brewing 2G stars (e.g., \citealt{bastian13}). So, special conditions  encountered only in the early Universe appear to be instrumental for the occurrence of the GC multiple population phenomenon.

\smallskip\noindent
\item
{\bf Ubiquity.} In almost all GCs studied in sufficient photometric and/or spectroscopic detail, evidence has been uncovered for the presence of 2G stars (in particular in the unprecedented sample explored by our Legacy Survey). This suggests  that the production of multiple populations is an unescapable outcome of the very formation process of GCs. None of the proposed scenarios is in obvious tension with this constraint, hence ubiquity is not considered any further in this paper. However, it remains essential for any GC formation theory to account for the formation of multiple stellar generations as an almost inescapable outcome, as opposed to a fortuitous event taking place only in a few cases and under special circumstances.

\smallskip\noindent
\item
{\bf Variety.} While virtually all GCs harbor  multiple populations, no two  GCs are alike, i.e., each cluster  has its own specific pattern of multiple populations, ranging from a minimum of two, up to seven and possibly more, each with specific chemical composition. Thus, there must be large cluster-to-cluster differences in the way in which materials from the 1G donors are eventually incorporated into 2G stars and/or in the chemical composition of such materials.

\smallskip\noindent
\item
{\bf Predominance.} 2G stars are not a minor component in most GCs, but may even dominate especially in the central regions where their fraction can largely exceed $\sim 50\%$, see in particular Paper II and Paper III. For example, in the central regions of NGC 2808 the putative 1G
makes up only $\sim 23\%$ of the cluster population (Paper III).

\smallskip\noindent
\item
{\bf Discreteness.} A major characteristic of multiple populations is that within each cluster they can be separated into quite
{\it distinct} sequences in various colorÐ-magnitude diagrams (CMD) and/or in appropriate two-color plots, as opposed to a continuous {\it spread}. To be sure, for quite some time spectroscopic evidence has instead been consistent with a continuous spread in chemical compositions, particularly suggestive in the case
of the ubiquitous O-Na anticorrelation (e.g., \citealt{carretta09}). However, the apparent continuity of spectroscopic abundances has been widely suspected of being due to measurement errors blurring an underlying discreteness, which is so evident in photometric data (e.g., \citealt{renzini13}). Quite recently, \cite{carretta14} was indeed able to show that in the case of NGC 2808, with more accurate abundances,  the earlier continuous O-Na anticorrelation indeed splits into at least three separate clumps. (For a previous example of spectroscopic discreteness see \citealt{marino08}) Discreteness is quite powerful at discriminating among competing scenarios and therefore we expand on it further in Appendix A.

\smallskip\noindent
\item
{\bf Supernova avoidance.} In most GCs the various multiple (2G) populations share the same metallicity [Fe/H] with the primary (1G) population, and do so within better than $\sim 0.1$ dex. Exceptions are the well known cases of $\omega$ Cen and Terzan 5,  but less extreme metallicity differences are now documented for quite a few other clusters such as M2,  M22,  M54, NGC 185, NGC 5286  and NGC 5824  (see Paper I and \citealt{marino11}, for references). Still, even in the case of $\omega$ Cen to enrich 2G stars to their observed iron content it is sufficient that only $\sim 2\%$  of the iron ejecta of the core-collapse supernovae from the 1G were retained and incorporated into 2G stars, while 98\% of such ejecta were lost by the system \citep{renzini13}.
Indeed, the main metal rich component of this cluster contains just $\sim 25\,\msun$ of additional iron compared to the iron content of the 1G component \citep{renzini08}. The core collapse supernovae from the 1G population of $\sim 2\times 10^6\,\msun$ at origin  have instead produced $\sim 1,000\,\msun$ of iron, with most having been ejected and only $\sim 2\%$ retained in 2G stars. This estimate is based on the present mass of the 1G component, so it is only an upper limit; if the 1G was substantially more massive at its origin, a smaller fraction would suffice.  Thus, in all GCs, 2G stars have experienced very
little contamination by supernova products, or none at all, with  only a very small fraction of such products having been incorporated into 2G stars. Every 
GC-formation scenario  needs to account for such avoidance of supernova ejecta. 

\smallskip\noindent
\item
{\bf Hot CNO and NeNa processing.}
A distinctive characteristic of 2G stars is the chemical composition
that results from CNO-cycling and {\it p}-capture processes at high
temperatures. All candidate 1G donors have indeed been chosen
purposely because they are a site of such processes. Thus, every
scenario should quantitatively account for the variety of composition
patterns exhibited by 2G stars in all GCs that have been
studied. Here, however, each scenario depends on the specific
nucleosynthesis yields of the invoked 1G donors, which all are
extremely (stellar-) model-dependent. For example, theoretical AGB
yields differ dramatically from one set of AGB models to another:\
some sodium is produced by the AGB models of Ventura et al.\ (2013)
but it is instead mostly destroyed in the models of Doherty et al.\
(2014), and comparable uncertainties are likely to affect the yields
of the other candidate polluters. Actually, \cite{bastian15}
argue that none of the yields used in each of the four scenarios is
able to reproduce the observed abundance patterns, even by appealing
to an {\it ad hoc} dilution with pristine material. This does not
necessarily invalidate any of the scenarios; instead it clearly calls
into question the stellar models used to calculate those yields. It is
important to realize that such theoretical yields depend on several
parameters needed to describe bulk motions of matter inside stars (e.g.,
rotation, convection, mixing, mass loss) and that only a tiny fraction
of this parameter space has been explored so far \citep{renzini14}. For these reasons we believe that at this
stage arguments based uniquely on the chemical composition of 2G stars
cannot be conclusive, either for or against any of the proposed
scenarios.

\smallskip\noindent
\item
{\bf Helium enrichment.} 
The discovery of the helium-rich 2G stars, first in $\omega$ Cen and in NGC 2808, and then in virtually all other GCs that have been studied,
has been bewildering and has completely changed our view of GCs and
their formation (see references in the previous papers of this
series). The ubiquity of the helium enrichment in 2G stars is extensively documented by the multi-band  database provided by our Legacy Survey.
All invoked 1G donors  do meet the requirement of shedding
helium-enriched material out of which 2G stars may form, though they may predict different yields of helium  relative to other elements, most notably oxygen and sodium.

\smallskip\noindent
\item
{\bf Mass Budget.}
The mentioned predominance of the 2G component in many GCs is a  challenge for all possible scenarios, as only a small fraction of the initial 1G mass is delivered
with the composition required for 2G stars. For example, with a
Kroupa/Chabrier-like  IMF the $\sim 3-10\,\msun$ AGB stars deliver only $\sim 10\%$ of the initial total mass of the first  generation. A way out from this difficulty is to postulate that the progenitors of today GCs were substantially more massive than their present-day GC progeny, with a strict lower limit of a factor of $\sim 5$ and possibly as much as a factor of $\sim 20$ or more. This is to say that GC progenitors would have lost at least $\sim 80-90\%$ of their mass before delivering the {\it naked} clusters as we see them today.  Especially demanding is the presence of oxygen depleted populations, which imply that such stars formed from material that was fully processed through 1G stars.
For example, in the central field of NGC 2808 the oxygen depleted populations account for $\sim 50\%$ of the total mass. Thus, depletion is more demanding for the mass budget than pollution, as e.g., a high sodium enhancement could in principle be achieved by polluting pristine gas with a relatively small mass of material highly enriched in sodium.
\end{itemize}

\section{Checking GC Formation Scenarios}

In this section we first sketch the four scenarios that have been
mentioned and then confront each of them with each of the binding
constraints, checking whether they can fulfill them as they are, or
only with major modification in some of the ingredients (e.g., the
stellar yields), or whether by their nature there is no way they could
satisfy  some  constraint.  These three different grades of
(im)plausibility are referred to in Table 1 as "OK", "TBD", and
"Nix", respectively. The four scenarios are now examined in detail, in order of
decreasing mass of the 1G donor. In doing so we will not discuss
whether a particular scenario is physically
plausible on general grounds, e.g., whether or not a SMS can form, or whether low-mass
stars can form in an extruding disk, etc., even if some of these
scenarios remain highly conjectural at this stage. As stated, we limit
ourselves to checking whether the binding constraints can be
fulfilled.

\subsection{Supermassive Stars}
 \cite{denissenkov14} and \cite{denissenkov15} build on the idea that within a young GC the most massive stars will sink to the center by dynamical friction and coalesce there, hence forming a SMS of 
 $\sim 10^4\,\msun$. An object of this kind would be fully convective, with a luminosity close to or exceeding the Eddington luminosity, hence  would lose mass at high rate. Since full convection makes the SMS chemically homogeneous, as it evolves its wind would be progressively enriched in helium and of products of CNO cycling and $p$-capture reactions: the kind of
composition one wants for 2G stars. This scenario
 quite naturally fulfills the ``GC specific" requirement, as the GC environment is indeed instrumental for the formation of the central SMS.  Variety may be
expected, e.g., from different possible masses of the SMS and from
different timing of the 2G star formation. Discreteness of the multiple populations may be accommodated appealing to separate bursts of star formation, that take place at different stages of pollution of the ISM by the SMS. Supernova avoidance is much more difficult to accommodate in this scenario, because the SMS and 2G stars should both form  before a fraction of a percent of massive stars  could explode. Because of this contrived timing, we give a Nix for this constraint.
Denissenkov \& Hartwick  want the SMS to be in a very precise range around $\sim 10^4\,\msun$, and should be ``.. neither $\sim 10^3\,\msun$ nor $\gsim 10^5\,\msun$",  otherwise their central temperatures would be either too low or too high to produce $p$-capture elements in the right mix. In \cite{denissenkov15} the SMS mass is chosen to be $7\times 10^4\,\msun$, because its central temperature of $\sim 75\times 10^6$ K ensures a sizable oxygen (and magnesium)  depletion without destroying sodium. Moreover, it is assumed that the SMS fragments and falls apart when helium has increased to $Y=0.4$, the upper limit of the helium abundance in 2G stars. This scenario faces an insurmountable difficulty   with the mass budget constraint:
for example, the major 2G population in $\omega$ Cen has a mass of $\sim 10^6\,\msun$ and has been enriched in helium by $\Delta Y\sim 0.1$ \citep{piotto05}, for which $\sim 10^5\,\msun$ of fresh helium is  required. So, the limit imposed to the mass of the SMS makes it impossible for them to deliver the required amount of helium and we give a Nix to both the mass budget and the helium constraints.
Moreover, the helium-rich 2G stars are also highly depleted in oxygen, implying that all their material was previously processed through 1G donor stars. Again, the same argument for the excess helium can be spent for the oxygen depletion: SMSs do produce helium and destroy oxygen, but a few $\times 10^4\,\msun$ SMS is simply not massive enough to produce helium enhancements and oxygen depletions on the scale that is observed in 2G stars in GCs. Appealing to multiple SMSs would not work either, as by the same token dynamical friction would force them to coalesce even faster than 
massive stars would do.

\subsection{Fast Rotating Massive Stars}
In the current version of this scenario \citep{krause13}, 2G stars form within the extruding disks of FRMS, with or without dilution with pristine materials that are supposed to exist in their vicinity. As such, if the process exists in nature, it is not specific
to GCs but to all FRMSs, whether in a GC or not. So if this process
worked, the result would be that stars with the chemical patters of 2G
stars in GCs would be nearly as frequent in the Galactic-halo field as
in GCs, which is at variance with the observations. Hence we give a
Nix for GC specificity. Given the star-by-star formation process, one
may expect little variation from one batch of 2G stars to another, as
2Gs would arise from the contribution of very many FRMSs; but perhaps
the scenario could be tuned to do that, so we assign a TBD to the
variety constraint.

Discreteness of 2G populations is, on the other hand, an
insurmountable difficulty for this scenario, since \cite{krause13} admit that "a consequence of this way of star formation is that
the distribution of abundances will always be continuous."  Supernova
avoidance is also a major problem. The fast winds from massive stars,
and from their supernova explosions, make the extremely crowded
central regions of GCs too harsh an environment for extruding disks to
avoid contamination by supernova products or even for them to survive
at all. The mass budget and detailed abundances of 2Gs are also a problem
for this scenario, as for all others, but as in the case of the other
1G donors, we cannot exclude the possibility that the actual yields
may be different from those computed so far and that the mass-budget
problem could thus be solved. Hence these last constraints do not
result in a fatal argument against this scenario.

\subsection{Massive Interacting Binaries} 
This scenario was first proposed by de Mink et al.\ (2009) and has been
further elaborated by \cite{bastian13}. In MIBs the forced
rotation of the primary envelope would cause mixing, which, if
reaching down to the hydrogen-burning shell, would result in CNO and
$p$-capture processing of the whole envelope, hence leading to helium
enhancement, oxygen depletion, etc. The processed envelope would then
be shed in a subsequent common-envelope phase of the MIB, thereby
going to replenish the ISM. In the original de Mink et al. scenario 2G stars would then form
out of  this material.
In the \cite{bastian13} version, MIB
ejecta would instead be swept up by the circumstellar disks of young,
low-mass stars and eventually dumped onto the stars themselves. The
scenario is certainly GC-specific, as the high density of a GC is
needed to ensure a sufficiently high density of the ISM to be built up
for appreciable accretion to take place.

Variety, i.e., large cluster to cluster differences, might be accommodated  in this scenario, that instead encounters insurmountable difficulties in producing multiple GC populations that are discrete. The problem is that large yet continuous star-to-star differences in the amount of swept/accreted material are naturally
expected in this scenario, while there is no known mechanism that would lead to quantized accretions resulting in the sharp discreteness exhibited by the composition of 2G stars in all GCs, with the extreme cases like $\omega$ Cen, M2, and NGC 2808, with their more than five distinct populations. Moreover, there is no way to establish and maintain the observed 1G/2G dichotomy, because the surviving 1G should  have avoided accretion completely, while the 2G stars would have been  dominated by it, with no intermediate cases (but see Appendix A for still a possible role for accretion).

Dichotomy/discreteness is indeed a killing argument for any accretion
scenario for the origin of multiple GC populations (e.g., \citealt{renzini13}). Supernova avoidance is another insurmountable difficulty for
this scenario too  (and for its original de Mink et al. version as well), as the MIB 1G donors would inevitably coexist with
the supernovae from single stars as well as from MIBs themselves, so the ISM
could not be made exclusively of the preferred MIB common-envelope
ejecta. \cite{bastian15} argue that the available yields plus
dilution cannot account for the detailed composition patterns that are
exhibited by GC multiple populations, but we regard this as a minor
concern compared with those just mentioned.  The mass-budget problem
may afflict this scenario as it does the others, but again this is not
necessarily a fatal flaw.

\subsection{AGB Stars} 
AGB stars have long been entertained as the possible origin, first of GC composition {\it anomalies} (e.g., \citealt{dantona83,renzini83,iben84}) and then of the composition  patterns of GC multiple populations (e.g., D'Ercole et al. 2010 and references therein). Indeed, during their AGB phase stars in the mass range $\sim 3-4\,\msun$  to $\sim 8\,\msun$ experience the so called hot bottom burning (HBB) process (see e.g., \citealt{ventura13},  whereby very high  temperatures are reached at the bottom of the convective envelope, thus allowing efficient $p$-capture nuclear processing. This mass range should be regarded as  indicative, because it is somewhat model dependent, e.g., on whether overshooting from convective cores is assumed and on its extension. Below $\sim 3\,\msun$ the HBB process does not operate and the AGB is populated by carbon stars. Since none of the 2Gs so far discovered is made of carbon stars one is forced to conclude that in this scenario the 2G formation must be completed before $\lsim 3\,\msun$ stars evolve into the AGB phase. It is important to note that the maximum temperature reached at the base of the convective envelope is a strong function of stellar mass, e.g., in the low-metallicity models of Ventura et al. (2013) it increases from 
$\sim 80\times 10^6$ K in $3\,\msun$ models to $\sim 140\times 10^6$ K in $7.5\,\msun$ models. Moreover, these values are extremely sensitive to the treatment of envelope convection, which remains a major source of uncertainty. The mentioned  mass range may be extended up to $\sim 10\,\msun$ by including among the 1G donors the so-called super-AGB stars, i.e., stars that ignite carbon in the core, shed large amount of helium and CNO-processed material, and die either as ONeMg white dwarfs or electron-capture supernovae (e.g., \citealt{ritossa96}). Thus, in this scenario the fact that stars in a wide range of stellar masses contribute the material to form 2G stars ensures that   $p$-processing takes place over a wide range of temperatures, with e.g., sodium being mostly provided by lower mass stars and oxygen-depleted material by the more massive ones.

In the specific model of \cite{dercole10} a massive GC progenitor  has a major episode of star formation leading to the first (1G) generation, but is then emptied of residual gas by supernova feedback. At the end of the supernova era, AGB (and Super-AGB) ejecta start to accumulate within the potential well of the system, since their ejection velocity ($\sim 10$ km s$^{-1}$)  is lower than the escape velocity. A cooling flow is then
established, leading to accumulation of gas within the original nucleus, until one or more starbursts make the 2G stars. Dilution with pristine gas was also invoked in the attempt to better reproduce the observed O-Na anticorrelation. In the subsequent dynamical evolution of the system most of 1G stars would be lost e.g.,  via
tidal interactions with the MW (or parent) galaxy, leaving a {\it naked} GC with comparable fractions of 1G and 2G stars. We consider here only the main features of this scenario, rather than its specific  incarnation in a particular model. For example, rather than a more massive, compact GC  the progenitor  may have been a nucleated dwarf galaxy (as suggested by e.g., \citealt{bekki06}),

GC-specificity, variety, and discreteness are probably not a problem for this scenario. The deep potential well of the progenitor is instrumental in retaining the AGB ejecta, thus allowing subsequent star formation events to take place from this material. In the GC progenitor the chemical composition of the ISM is rapidly changing as
it gets replenished by the ejecta of AGB stars of progressively lower initial mass; hence each star-formation burst will have a specific composition. Thus, discreteness arises from bursts of 2G star formation being separated by periods of (relative) inactivity. Variety in different GCs can arise from different numbers of bursts and their different timings. Thus, the intimate stocharsticity of star formation would  ensure variety. The AGB era during which 2G stars would form is indeed bound in time by the end of the core-collapse supernovae on one side and by the ultimate gas removal (by either Type Ia supernovae or interaction with the environment) on the other. This time interval, of the order of $\sim 10^8$ years, may well change from one cluster to another, depending on the detailed star formation history that led to the first generation. Similar arguments can be applied to other scenarios as well, where 2G stars are produced in star formation events.

Avoidance of contamination by 1G core-collapse supernovae is automatically fulfilled, but if there is more than one 2G star-burst, supernovae from stars of the first 2G
burst may contaminate the later-formed generations, an aspect that is further discussed in a following section. The postulated dilution with
pristine gas remains problematical: where was such material stored in the meantime, and how did it avoid being contaminated by 1G
supernovae? These questions remain unanswered.

The mass budget is a problem that is {\it solved} by appealing to a sufficiently massive precursor. Quantitative estimates of how massive this precursor needs to be 
depend on several uncertain factors such as  the mass range of the 1G AGB donors, the star formation
efficiency, the 2G IMF and the incidence of the postulated dilution.  Particularly critical is the star formation efficiency,  i.e., the fraction of the available gas from 1G ejecta  that is effectively turned into 2G stars (e.g., \citealt{renzini13}).  To minimize the mass budget D'Ercole et al. assume that the IMF of 2G stars is truncated at a value $\lsim 8\,\msun$, which also ensures supernova avoidance from one 2G starburst to the next.  Moreover, the absence of 2G supernovae would prevent gas that is not used by a 2G star-formation episode from being expelled from the
system, hence keeping it available for subsequent star-formation events. Thus,  even if the star formation efficiency of individual
bursts is low (say $\sim 10-20\%$), bursts of star formation would continue
until virtually all of the available gas is turned into 2G stars. In such a way the mass required for the progenitors could
be reduced  perhaps optimistically) to $\sim 5-10$ times the present mass of a GC.  Yet, the nature of GC progenitors remains a critical problem for this as for all other scenarios. In Section 4.1 we further address this issue.

\begin{table*}
\centering
{
\caption{The cross-check of the four scenarios  discussed in the text vs the observational constraints set by the properties of the second generations.}
\vspace{5 truemm}
\begin{tabular}{lccccccc}
\hline
\hline
 Scenario &  GC Specific & Variety & Discreteness  &  SN Avoidance & Mass budget & Hot p-captures & Helium\\
\hline
SMS& OK& OK & TBD  & Nix & Nix & Nix & Nix\\
FRMS & Nix& TBD & Nix& Nix& TBD & TBD& TBD\\
MIB$_{\rm Acc}$ &  OK& OK & Nix & Nix & TBD & TBD & TBD\\
MIB$_{\rm SF}$  &  OK& OK & OK & Nix & TBD & TBD & TBD\\
AGB&   OK& OK & OK & OK & TBD & TBD & TBD\\
\hline
\hline
\end{tabular}
}
\label{tab:crosscheck}
\end{table*}

One enduring problem with AGB stars as 1G donors  is that they tend to produce a {\it correlation} between oxygen and sodium, rather than an {\it anticorrelation}, as actually observed. (See e.g., Figure 3 in \citealt{dantona11}.) Assuming that the mass range of AGB producers of $p$-capture elements is somewhere
between $\sim 3$ and $\sim 8-10\,\msun$, towards the low mass end of this range these AGB models shed material that is both O rich and Na rich. At the opposite mass end they shed material that is both O- and Na-depleted. 

This makes it difficult to match the observed anticorrelation, especially
in cases of  extreme oxygen depletion. However, AGB yields are
extremely sensitive to several, interlaced parameters describing processes
such as convection, mixing and mass loss. So, 
future calculations may deliver yields more compatible with the survival of
sodium in matter processed by HBB at high temperatures.
In Section 4 we recall what are the main physical processes affecting the AGB yields and we speculate on how AGB models might  be tuned to produce yields in better
agreement with the observational requirements. All in all, there appears to be no blatant show-stopper for this scenario and, as
reported in Table 1, no Nix is assigned to it.

\subsection{Multiple Stellar Populations or Multiple Stellar Generations?}
In all scenarios discussed above, except one, more than one star formation event takes place, a first one out of pristine  material (1G) and one or more subsequent  events  where 2G stars  form out of material processed by  1G stars. The exception is the case of the scenario where ejecta from massive interacting binaries are accreted by the circumstellar disks of lower mass stars. In such case there is indeed only one star formation episode, i.e., only one stellar generation (1G), but some of the 1G stars are polluted by the ejecta of more massive stars of the same generation. However, by its very nature any accretion scenario is incapable of producing distinct, multiple stellar populations within individual GCs such as those being documented by the present Legacy Survey. Thus, we consider GC ``multiple  stellar populations" and ``multiple stellar generations" as synonyms and may use the two expressions interchangeably.  

\subsection{Summary of the Cross check of Scenarios and Constraints}
Table 1 summarizes the results of these cross checks. In the table we distinguish the two versions of the MIB scenario, one with circumstellar disk accretion (MIB$_{\rm Acc}$) and another with possibly discrete events of star formation (MIB$_{\rm SF}$).  Three scenarios,
namely SMS, FRMS and MIB$_{\rm Acc}$, fail to meet  two or more constraints and do
so because of their intrinsic nature, i.e., such failure does not
appear curable by fine tuning parameters. Supernova avoidance and
discreteness are the constraints which are more widely violated, and
to which insufficient attention, or none, has been dedicated by the
proponents of such scenarios. The MIB$_{\rm SF}$ option clearly violates only the supernova avoidance constraint.
The AGB is the only scenario that does
not appear to irreparably violate the seven constraints. Yet, it is
still far from providing an adequate, quantitative account for the
specific composition patterns so far documented in the first papers of
this series and in the references therein.  The mass budget problem remains, along with the still unknown nature of the GC precursors, i.e. the systems that nursed GCs as we eventually see them today. Whether the AGB scenario
could be upgraded to meet the observed patterns quantitatively is
addressed in the next section.

\section{Saving the AGB Option?}
The {\it binding constraints} from our Legacy Survey are sufficient to clearly falsify three of the four examined scenarios, but still do not provide fatal evidence against the AGB scenario.
Yet, even if we did not assign any "Nix"  to the AGB option, this scenario encounters if not fatal, at least severe difficulties in accounting for all the accumulated evidence on the multiple populations of GCs. The mass budget is one, the detailed chemical composition of 2G stars is another.  In this section we speculate on whether possible ways may exist of  upgrading this scenario to better match the binding constraints. We emphasize that we do so for a lack, at least temporarily,  of any better alternative.

\subsection{The Mass Budget and the IMF of 2G Stars}
A possible solution (or alleviation) of the mass budget problem has
already been mentioned, i.e., the \cite{dercole10} postulate
of a different IMF between 1G and 2G stars, with that of 2Gs being
truncated at a mass close to or below   $\sim 8\,\msun$.  This {\it ansatz} has three beneficial effects, it reduces the mass budget directly as well as indirectly (by allowing a virtually $\sim 100\%$ star formation efficiency) and avoids supernova pollution from one 2G to another. Yet it remains unproved. 

One possible reason for a different IMF for 2G stars comes from the fact that such stars have to form in an environment already densely occupied by 1G stars.
The typical central density of a massive GC  is $\sim 10^5\,\msun$ pc$^{-3}$, corresponding to a number
density of atoms $n\simeq 10^7$ cm$^{-3}$.  Moreover, 1G and 2G stars have  comparable number density in a GC central region and often the 2G even prevails.
With a Chabrier/Kroupa IMF,  $\sim 150$ stars (more massive than $0.1\,\msun$) are formed every 100 $\msun$  of gas that goes  into stars, hence central densities exceed  $10^5$
stars per cubic parsec (and might have been even higher at formation time). Thus, 2G star formation takes place in an environment already inhabited by an extremely dense stellar system. This may well be a different
mode of star formation, compared to the case of a molecular cloud virtually devoid of pre-existing stars (Renzini 2013).  To our knowledge, star formation in an extremely densely  populated stellar system is a mode of star formation never explored so far. Yet, this must be the mode to form 2G stars in virtually all GCs.  For the time being, we can say that if there is a situation in which star formation takes place with a different IMF, this may well be the core of a proto-GC.  Might massive star formation be inhibited in such environment? This does not appear to be the case in the vicinity of the Galactic center, where stellar densities may be even higher, and where massive stars appear to have formed, either by coalescence of less massive ones or in counter-rotating disks (Genzel et al. 2003), though conditions in the Galactic center may not be representative of those in proto-GCs.

 Besides helping with the Na-O anticorrelation, the postulated dilution with pristine material also has  the beneficial effect of somewhat alleviating the mass budget issue. However, its origin remains problematical along with its supernova avoidance. One less controversial form of dilution which must occur to some extent is via the ejecta of common-envelope binaries of intermediate mass stars, as advocated by \cite{vanbeveren12}. 

How over $\sim 80\%$ of the progenitor mass would have been removed remains problematical. There appears to be no correlation of the 2G/1G ratio with galactocentric distance \citep{bastianl15}, confirmed by the larger and homogeneous dataset of our Legacy survey (Milone et al. in preparation). This may argue against the progenitor being a compact, just more massive GC and may favor the nucleated dwarf galaxy option, with the less bound body of such an object being more easily stripped. 

\cite{larsen12} have argued that the Fornax dSph galaxy with its own GCs may set an upper limit to the possible mass budget. Based on two O-poor/Na-rich stars 
found in two among the four metal poor GCs (from \citealt{letarte06}) it is inferred that also Fornax  GCs harbor multiple stellar populations. Larsen et al. note that the metal poor component of this galaxy is only $\sim 4-5$ times more massive than its GCs together, hence no more donor mass than 4--5 times the
present mass of the GCs would have been available for the production
of their 2G stars. However, it remains to be seen  what was the original mass of this dSph galaxy. Indeed, its specific frequency of GCs  is the absolute highest known, i.e., $\sim 26$  times  that of the MW galaxy, which suggests that also Fornax  may have lost a significant fraction of its original stellar mass. This may be even more the case for its metal poor component, given that the specific GC frequency jumps to $\sim 400$ if the calculation is restricted to only the metal poor component. So, we regard as interesting but not yet compelling the proposed upper limit for the mass budget.

\begin{figure}
   \centering
\includegraphics[width=\columnwidth]{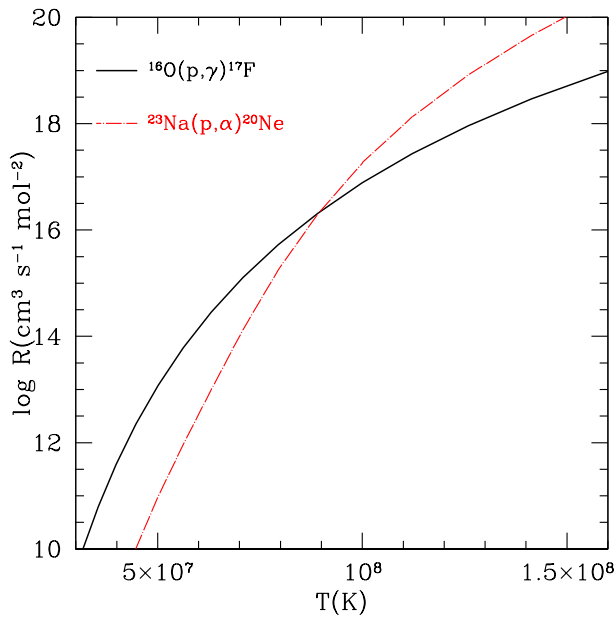}
\caption{The rates of the  $^{16}$O$(p,\gamma)^{17}$F and $^{23}$Na$(p,\alpha)^{20}$Ne reactions as a function of temperature, showing that for 
$T \lesssim 10^8$   
K oxygen is destroyed faster than sodium, whereas sodium is destroyed faster above this temperature.}
 \label{onarates}
\end{figure} 

\subsection{Sodium-Oxygen Anticorrelation, and the like}

As mentioned in Section 3.4, the main difficulty encountered by the AGB scenario consists in predicting the observed chemical patterns observed in 2G stars, most notably the oxygen-sodium anticorrelation. Thus, it is  worth expanding a bit on the physics of why this happens. Upon arrival on the AGB, stars in this mass range are only slightly depleted in oxygen and somewhat enriched in sodium. They are also slightly depleted in carbon and enriched in nitrogen and helium. All this is due essentially to the second dredge-up (2DU), when envelope convection penetrates through the extinguished hydrogen shell into helium layers that were processed by the hydrogen-burning shell during the previous evolutionary phases. Indeed, in this shell oxygen was highly depleted in favor of nitrogen while sodium had been produced by the $^{22}$Ne($p,\gamma)^{23}$Na reaction, with $^{22}$Ne being also replenished by two successive $p$-captures on $^{20}$Ne. Since initially $^{22}$Ne is more abundant than $^{23}$Na and $^{20}$Ne is $\sim 100$ times more abundant, even a relatively small reduction in the abundance of Ne isotopes can result in a factor of $\sim 10$ increase in the 
surface abundance of $^{23}$Na. Upon settling on the AGB, these stars start experiencing the HBB process and now the whole convective  envelope works as a reservoir of neon isotopes ready to be turned into $^{23}$Na. Thus, brought by convection to the HBB layers, part of these neon isotopes are then converted to $^{23}$Na, further enhancing the surface abundance of sodium (see e.g., Figure 6 in \citealt{ventura05a}). However,  as the sodium abundance increases, so does the rate at which it is destroyed by its own $p$-captures, via
the reactions $^{23}$Na$(p,\gamma)^{24}$Mg and $^{23}$Na$(p,\alpha)^{20}$Ne. At the same time oxygen is being destroyed by the reaction $^{16}$O$(p,\gamma)^{17}$F and subsequent reactions, eventually turning  this oxygen into nitrogen.  So, after an initial spike  of $^{23}$Na production, both sodium and oxygen tend to be destroyed, which is why it happens that way in,  e.g., the AGB models of \cite{dantona11}  and \cite{doherty14}.

Therefore, what matters is ultimately the relative rate at which oxygen and sodium are destroyed and on the timing in the interruption of these processes when the envelope is eventually lost in a (super)wind and the Post-AGB phase begins. The existence of oxygen-poor and sodium-rich stars among  the 2G stars of many GCs argues  for oxygen being destroyed faster than sodium. How could this happen?  Figure 1 shows the rates of the $^{16}$O$(p,\gamma)^{17}$F and $^{23}$Na$(p,\alpha)^{20}$Ne reactions as a function of temperature (this latter being the dominant channel for $p$-captures on $^{23}$Na). Thus, at lower temperatures  ($T\lsim 10^8$ K) oxygen is destroyed faster than sodium: a result of the lower tunneling probability through the Coulomb barrier of sodium compared to that of oxygen.
However, at higher temperatures the difference in tunneling probabilities decreases and what becomes dominant is the fully nuclear part of the cross section: the destruction of sodium becomes faster than that of oxygen because the destruction reaction is mediated by the {\it strong} interactions (revealed by the emission of an $\alpha$ particle) whereas the oxygen destruction reaction is mediated by the much weaker {\it electromagnetic} interactions (as revealed by the emission of a $\gamma$ photon). Thus, above $\sim 10^8$ K sodium is destroyed faster than oxygen. By the same token, it is clear that the AGB evolution should end before this Ne-Na cycle has reached equilibrium, i.e., near equality between production and destruction rates of neon and sodium isotopes. For, equilibrium disfavors sodium as it is produced by {\it electromagnetic} nuclear reactions and destroyed by {\it strong} nuclear reactions.

This means that if one wants to produce AGB yields that are oxygen depleted and still sodium rich, then   the HBB should work at temperatures below $\sim 10^8$ K
in a suitable fraction of the AGB stars. Incidentally, this is precisely why \cite{denissenkov15}  want their SMSs to work at $\sim 75\times 10^6$ K.  Actually, the AGB scenario  offers an important opportunity, in that the AGB yields result from the contribution of stars in and extended mass range (within roughly $\sim 3$ to $\sim 10\,\msun$), hence with an extended range of temperatures at which the HBB process has operated, because such temperature is a strong function of stellar mass, metallicity and changes in the course of the evolution of each individual AGB star  (e.g., Renzini \& Voli 1981). In the most massive AGB stars the HBB temperature may indeed be so high that  even the abundance of Mg, Al, Si, and K can be affected by $p$-capture reactions, and indeed demanded by the observed abundances of these elements in the 2G stars of some GCs (e.g., \citealt{carretta09,carretta14, cohen12, mucciarelli12}).

This is illustrated in Figure 2, where the temperatures above which the various $p$-capture reactions effectively operate is shown, including those producing Al, Si and K at the expenses of respectively $^{24}$Mg, $^{27}$Al and $^{38}$Ar (Ventura et al. 2011, 2012), an opportunity that only the AGB scenario can potentially offer. 
The figure shows that models of either MIBs or FRMSs operate at temperatures not exceeding $\sim 65\times 10^6$  K, so they can easily process oxygen to nitrogen and do not destroy sodium, but fail to appreciably turn magnesium into aluminum. 
The preferred central temperature for  the SMS models  is indicated by the blue vertical bar at $\sim 75\times 10^6$ K, sufficient to destroy O while preserving Na and ensuring some conversion of $^{24}$Mg into $^{25}$Al, but being too low to allow production of Si and K.
Finally, the HBB temperature range potentially covered by AGB stars extends from well below to somewhat above $\sim 10^8$  K, encompassing a wide variety of situations,
from oxygen depletion and sodium production, to  Al, Si and K production. However, stars producing these heavier
$p$-capture elements will necessarily destroy sodium, hence other, less massive AGB stars should produce it while still destroying oxygen. As mentioned above,
so far none of the incarnations of the AGB scenario has fully accounted for the detailed abundance patterns exhibited by 2G stars. 

We see two potential ways of {\it saving} this scenario, i.e., to make it to produce $p$-capture elements (and helium) in the observed proportions. 
The simplest way is to assume that the actual cross section of the $^{23}$Na$(p,\alpha)^{20}$Ne reaction is somewhat lower that the recommended value by Angulo et al. (1999). Indeed, with a factor of $\sim 5$ reduction the temperature above which Na is destroyed faster than O will increase from $\sim 90\times 10^6$  K  to over $\sim 115\times 10^6$  K, as illustrated in Figure 2. The selective reduction of just this cross section would suffice to establish a Na-O anticorrelation in better agreement with the observations, without affecting other successes of current AGB models, such as the $p$-capture production of Al, Si and K \citep{ventura12, ventura13}.
   \begin{figure}
   \centering
\includegraphics[width=\columnwidth]{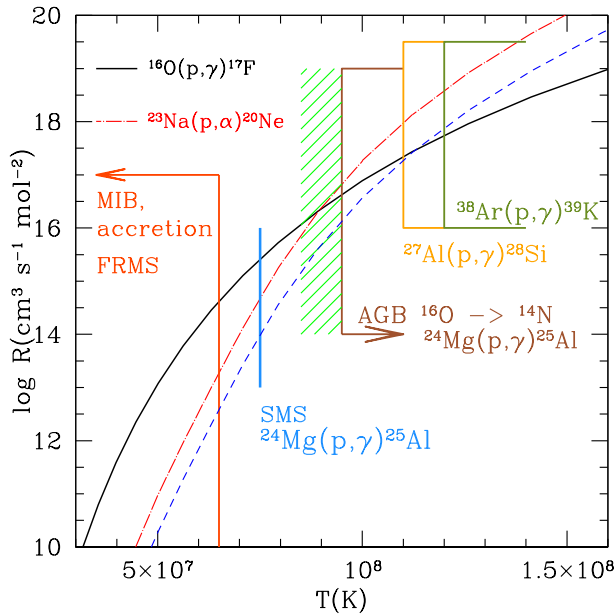}
\caption{As in Figure 1, the figure shows as a function of the temperature the rates of the two reactions which mainly determine the O/Na ratios in AGB stars. The rate of the $^{23}$Na burning is the lower limit allowed by the NACRE reaction rates as compiled by Angulo et al. (1999), 
i.e., a factor $\sim$2 below the recommended rate (red line) whereas the greed dashed line corresponds to a further reduction by a factor of 5. The favored temperatures by the various scenarios discussed in this paper (SMSs, FRMSs, MIBs and AGB stars) are also indicated as well as the temperatures above which the indicated reactions start to efficiently operate (see text). 
The green shaded temperature range corresponds to the maximum temperature in lower mass AGB models (e.g., Ventura et al. 2013),
where only marginal oxygen depletion may take place.
}
 \label{onarates2}
\end{figure}

The alternative, assuming current cross sections to be correct, would be to attribute the mismatch to insufficient exploration of the AGB parameter space, a rather laborious possibility to pursue. AGB models rely on assumptions concerning mass loss, mixing and superadiabatic convection, all poorly understood processes.
Therefore, it is not surprising if existing AGB models come tantalizingly close, but not quite enough,  to produce chemical yields that satisfy the 2G requirements. 
The residual mismatch could then be due to insufficiencies in the adopted parametrizations of these processes and/or in the parameter combinations so far explored.
In \cite{dantona11}, \cite{ventura13} and \cite{doherty14} AGB models, above $\sim 4-5\,\msun$,  the HBB process burns at   $T\gsim 10^8$ K, hence destroys both oxygen and sodium, whereas at lower masses sodium is produced but not much oxygen is destroyed. The  divide between these two mass ranges  depends strongly on the adopted efficiency  envelope convection: the more efficient convection, the higher the HBB temperature, hence  the lower the mass at which burning works at  $T\sim 10^8$ K  (e.g., \citealt{renzini81,ventura05b}). Existing AGB models  in which HBB burning operates  at $T\lsim 10^8$ K show a strong effect of the third dredge up (3DU), so that their C+N+O increases, oxygen is less depleted, or increases too, and sodium increases even too much (e.g., \citealt{fenner04}). Thus, acting solely on convective efficiency may help with sodium, but at the expense of worsening the match to e.g., the global CNO abundance.
A possible solution would require both to extend the AGB lifetime (e.g., by working on the assumptions made for the mass loss processes), in order to achieve a certain level of oxygen depletion, while at the same time  reduce the efficiency of mixing via the 3DU.  Indeed, the three parametrized processes, envelope convection (hence  efficiency of the HBB process for a given mass), 3DU mixing and mass loss, all influence each other in a closely entangled fashion \citep{renzini14}, and all together concur in determining the AGB lifetime, luminosity excursion and eventually chemical yields, all as a function of stellar mass. 
This is to say that the exploration of the AGB model parameter space is not a simple task and whether one can accommodate even Si and K production in the most massive AGB stars while still delivering Na-rich yields from the whole AGB mass range remains to be demonstrated.
 If this approach should succeed, the need for the postulated dilution with pristine material could  also be reassessed. Anyway, if so far AGB models have failed to fully reproduce the composition of GC 2Gs, then what we can do is turn the obstacle around, and see what GC 2Gs can tell us about the evolution of AGB stars, and use them as a guide to improve upon the construction of AGB models. Hopefully, if such an endeavor succeeds, we will gather a better understanding of both GC formation and of AGB evolution  -- which are perhaps one and the
same problem. 
%

\subsection{Oxygen Depletion without much Helium Enrichment}

A potential problem that at this point
in time appears to be common to all suggested polluters, i.e.,  that
large oxygen depletions are typically accompanied by a large helium
enhancement.  Instead, in some cases such as the GC 47 Tuc, a sizable depletion  in oxygen by at least a factor of $\sim 3$ \citep{carretta09} is accompanied by only a modest $\Delta Y\simeq 0.03$ increase in helium abundance \citep{milone12}.  
Bastian et al. (2015) have expanded on this issue, arguing that it is intrinsic to nuclear processing to have oxygen destruction and helium production to be closely correlated and they seem to believe that other than nuclear processes are responsible for the observed {\it anomaly}.  Before appealing to more exotic physics, again it may be worth exploring whether insufficiencies in the models or in their implementation  may have been responsible for the mismatch with the observations.

In this respect, it is worth recalling that most of the helium enrichment in AGB stars is due to the 2DU, which brings helium to the surface in  stars more massive than $\sim 3$ to 
$\sim 5\,\msun$, depending on composition \citep{becker79}. Oxygen is instead depleted via the HBB process, that operates in stars more massive than $\sim 3$ to $\sim 5\,\msun$, depending on composition and the assumed efficiency of envelope convection. In the case of all the other donors, CNO  processing occurs in the interior of hydrogen-burning main sequence  stars, so unavoidably it is  strictly linked to the helium production.
The helium  and O--Na yields of AGB stars are instead not so tightly bound to each other, as these elements  are  processed  at different times in the course of evolution (by the 2DU and HBB, respectively), and in a mass dependent fashion.  In principle, it is indeed possible that a mass range exists in which not much helium is brought to the surface by the 2DU whereas oxygen is significantly depleted by the HBB.  The amount of helium produced by the HBB is indeed quite small.  Of course, appealing to a reduced mass range of AGB donors would somewhat  exacerbate the mass budget problem.

\section{Summary and Conclusions}

In this paper we have first summarized the most salient properties of
multiple generations in globular clusters, as they have emerged from
the many observational studies of the last decade which have culminated with our {\it HST} Legacy Survey of Galactic Globular Clusters (Paper I). Such properties include GC Specificity, Ubiquity, Variety,
Discreteness, Supernova Avoidance, Hot CNO/NeNa Processing, Helium
Enrichment and Mass Budget.  Such observational evidence is used to
check whether four scenarios for the formation of multiple populations
are consistent with each of the constraints, or whether they violate
them in a possibly curable or incurable way.

The four scenarios differ for the nature of the stars producing the material used to form the second populations/generations of stars, namely supermassive stars (few $10^4\,\msun$, \citealt{denissenkov15}, fast rotating massive stars (25 to 120 $\msun$, \citealt{krause13}, massive interacting binaries ($\sim 25+15\,\msun$, \citealt{demink09} and AGB (+Super-AGB) stars ($\sim 3$ to $\sim 8-10\,\msun$, e.g., \citealt{dercole10}).  Our  cross-check examination (summarized in Table 1) indicates that the first three scenarios encounter unsurmountable difficulties in fulfilling one or more  of the above observational constraints, and we conclude that they are untenable (at least in their current form). Only the AGB scenario survives, but barely. 

The main difficulties encountered by  the AGB option concern the mass budget and  the detailed chemical composition of second generation stars. For the AGB scenario to work, the mass of the parent stellar system (dwarf nucleated galaxy? super star cluster?) needs to  have been at least $\sim 10$ times more massive than current GC survivors, i.e., with masses up to several $10^7\,\msun$ and possibly more. So, suitable proto-GCs remain to be firmly identified. For the chemical composition issue, the problem is that 2G stars which are depleted in
oxygen and enriched in sodium are rather common, whereas current AGB models deliver materials that if are depleted in oxygen then they are also depleted in 
sodium. This seem to us the most serious difficulty, hence we ignore minor ones such as the abundances of  lithium or s-process elements. 

One way of alleviating the mass budget problem is to assume that the 2G stars form with a different IMF compared to 1G stars \citep{dercole10}, in particular assuming that 2Gs consist only of stars less massive than $\sim 8\,\msun$. This would also allow supernova avoidance between one 2G and the next and could possibly allow a full  conversion of AGB ejecta into 2G stars with $\sim 100\%$ efficiency.  We emphasize that forming the 2G stars in an environment already extremely packed with 1G stars
(i.e., over $10^5$ stars per cubic parsec) is a star formation mode quite different from that prevailing under normal circumstances. We then speculate that such
an as-yet-unexplored mode of star formation may lead to an IMF which is devoid of massive stars. Even so, the nature of the GC progenitors and how they would have lost  most of their stellar mass  remain puzzling. The lack of a correlation of the 2G/1G ratio with galactocentric distance suggests that tidal stripping of a massive and compact progenitor may not solve the budget problem.

Concerning the chemistry, we note that the reason that current AGB
models fail may be that CNO and NeNa cycles operate in them at
temperatures above  $\sim 10^8$ K, when sodium is actually destroyed faster than oxygen. This discrepancy would be much alleviated if the cross section of the sodium-destroying reaction $^{23}$Na$(p,\alpha)^{20}$Ne were  actually a factor of a few lower than currently estimated, a possibility that future experiments may test. 
This simplest solution of the problem of the Na-O anticorrelation has the advantage of avoiding to jeopardize other successes of current AGB models, such as the extension to Al, Si and K of the involvement in $p$-capture processing, which requires temperatures well in excess of $10^8$ K.

Alternatively, we argue that tuning envelope convection to reduce somewhat below this limit the temperature at the base of the envelope --in a majority of AGB stars-- may  result in AGB chemical yields with low oxygen and still high sodium, hence in better agreement with the observations. Yet, this cannot be achieved without concurrently acting on the other parametrized processes, i.e., the third dredge up and mass loss, in such a way to ensure  a suitable extension of the AGB lifetimes while avoiding excessive CNO and s-process enhancements via the 3DU.  In other words, we suggest to put the AGB scenario under intensive care, and see whether 
it can be saved by an extensive exploration of the parameter space.  Success is by no means guaranteed.  If the AGB option were also to fail, we would be left without any viable scenario:  a new, totally different formation scenario for GCs and their
multiple populations would have to be invented. One may argue that one scenario does not necessarily exclude another, and that two or more 1G
donors may operate together. For the time being, however, we prefer to
avoid the intricacies that may arise in such a composite option.
Young massive clusters in the nearby galaxies show no signs of being forming 2G stars, whereas 2Gs are ubiquitous among our $\sim 12$ Gyr old GCs. Clearly, YMCs and GCs  may not form in the same way. This suggests that  special conditions prevailing only  in the early Universe may have been determinant in leading to GC formation with  their multiple generations.

What is really striking is the prolificacy with which Nature has made so
many complex systems, in contrast with our persistent inability to
understand how they formed and evolved to their present
state. Even the least implausible solution appears quite contrived and relies on several unproven assumptions.  
However, exploration of Galactic GCs has made great progress in
recent years, and the evidence is now fairly well documented, both
photometrically and spectroscopically.  Perhaps a new day is dawning,
with new opportunities, such as the spectroscopic follow-up of
photometrically selected sub-populations and of their spatial
distribution and kinematic differences within each cluster.  This is
what our team is setting out to pursue, starting with NGC 2808
(\citealt{bellini15b}, Marino et al., in preparation), and
continuing with a host of other GCs (see Paper I).  

The spatial distribution of the various sub-populations within each cluster
is not discussed in this paper,  because our {\it Legacy Survey} data pertain only to the central regions of the studied GCs.
Yet, in a few cases we know that radial gradients exist, with 2G stars being more centrally concentrated than 1G stars, e.g., in $\omega$ Cen \citep{sollima07}
and in 47 Tuc \citep{milone12}. Mapping the radial trends of the 2G/1G population ratios in most of the {\it Legacy Survey} clusters is an obvious next step
in the study of multiple populations and may add further critical constraints on formation scenarios.

There is also new territory to explore, concerning the GC populations
in other galaxies. This could tell us whether the multiple-population
phenomenon is also common within other GC families, and whether its
frequency depends on the nature of the host galaxy, thus giving us new
hints about, or constraints on, how GCs form. In this direction, 
\cite{bellini15a} have combined {\it HST} optical and UV data to
study almost 2,000 GCs in the core of the giant elliptical galaxy M87,
in the hope of finding  whether some of them may host
UV-bright multiple stellar generations.  Their experiment has reached
only partial success, but we believe that this kind of study can give
us important clues on GC formation and should be pursued.

Finally, a real breakthrough would be to catch GCs while they are
still forming (or shortly thereafter), i.e.,  at redshifts beyond 2 or 3,  at a lookback
time of 10--13 Gyr.  If their parent stellar systems were really as
massive as $10^7$ to $10^8\,\msun$, then their light should be
observable by JWST and by the next generation of extremely large
telescopes on the Earth's surface. Massive galaxies at these redshifts may well be encircled by a 
swarm of forming/young GCs which may not remain  below detection threshold for long.


\section*{Acknowledgments}
\noindent
APM acknowledges support by the Australian Research Council
through Discovery Early Career Researcher Award DE150101816.  AR, SC, and GP acknowledge
partial support by PRIN-INAF 2014, and GP acknowledges partial support
by "Progetto di Ateneo" (Universit'{a} di Padova) 2014.  AB and IK
acknowledge support from STScI grant GO-13297, provided by the Space
  Telescope Science Institute, which is operated by AURA, Inc., under
  NASA contract NAS 5-26555.

\vspace{0.8 truecm}
\noindent
{\bf Appendix: On Discreteness}
\\
To illustrate what we mean by {\it discreteness}, Figure 3 shows the composite multicolor plot for the red giants in  the GC NGC 2808, replicated from Paper III  where the indices  $\Delta_{\rm F336W, F438W}$  and \,$\Delta_{\rm F275W, F814W}$  are defined.  Similar plots, that we nickname {\it chromosomic maps}, are now being constructed for all 57 GCs in this {\it Legacy Survey} plus its pathfinders. We have chosen this particular cluster to illustrate the case, because it is one of those for which somewhat deeper integrations have been used, resulting in very small photometric errors ($\sim 0.01$ mag) that have allowed us to distinguish at least five  and possibly as many as seven distinct sub-populations. (A rigorous statistical estimate of the number of sub-populations just confirms what just an eye examination suggests.). While the presence of distinct sub-populations is evident, one important issue is whether each of them is a {\it simple stellar population}, i.e., whether all stars in each of the clumps in the chromosomic map have the same composition, or whether there is an intrinsic  dispersion  internal to each clump, possibly leading to marginal overlap between adjacent clumps. The reddening of the cluster is fairly high --$E(B-V)=0.24$-- hence differential reddening of order of $\sim 0.03$ mag could be expected. Although a differential reddening correction has been applied before constructing the indices shown in Figure 3 (see Paper III and references therein),  still we suspect  that errors in such corrections  are larger than pure photometric errors, resulting in combined errors of order of $\sim 0.02-0.03$ mag. 

This estimated error is still substantially smaller than the width of individual peaks in the histograms shown in Figure 2 and, unless there are other unaccounted sources of error, we suspect that individual clumps are not made of stars with identical composition, but a small dispersion exists among them. Such a dispersion can originate in two possible ways. One is that the individual bursts of star formation had a finite duration, hence stars formed at different phases of the burst were made of material with slightly different composition, as indeed the composition of the ISM was continuously changing being  the ISM continuously fed by AGB star of different mass.
Moreover, in between bursts star formation may have not vanished entirely, so one burst partially overlapped with the next one. Besides an intra-clump dispersion due to the detailed star formation history, accretion may add further dispersion, both on top of 1G and 2G stars. Indeed, if the Bondi formula applies one would expect 
appreciable accretion to take place during the first $\sim 10^8$ years (cf. Renzini 2013).

As documented in Paper III, Population A+B in Figure 3  represents the 1G of NGC 2808. Clearly, the elongated and clumpy distribution of this feature of the chromosomic map suggests that even the 1G stars are not chemically homogeneous, possibly by marginal contamination of some stars by supernova products
by 1G itself, or by accretion of AGB ejecta, or both.

   \begin{figure*}
   \centering
\includegraphics[width=1.7\columnwidth]{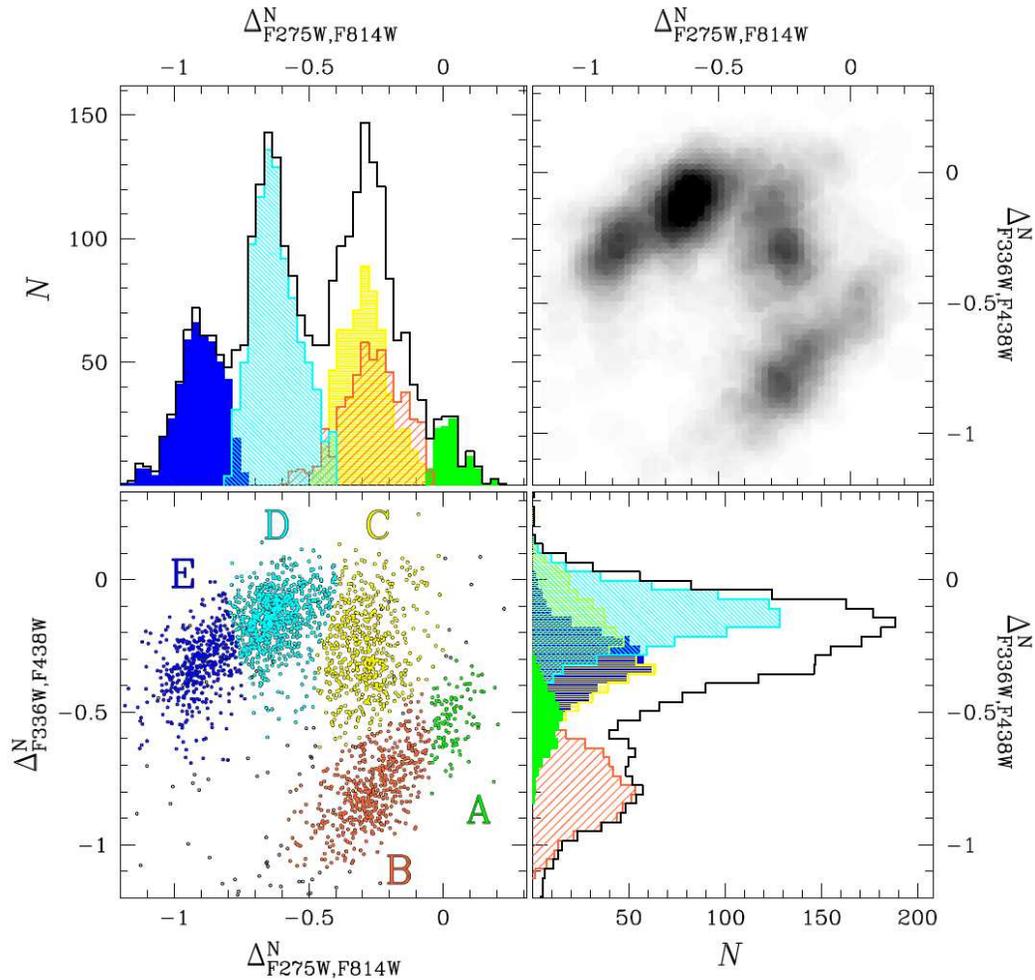}
      \caption{Reproduction from Milone et al (2015b) of the $\Delta_{\rm F336W, F438W}$ vs.\,$\Delta_{\rm F275W, F814W}$ diagram of NGC 2808. Stars in the A, B, C, D, and E groups are colored green, orange, yellow, cyan, and blue, respectively (lower-left).   The corresponding Hess diagram is plotted in the upper-right panel.  The histograms of the normalized $\Delta_{\rm F275W, F814W}$ and $\Delta_{\rm F336W, F438W}$ distributions for all the analyzed RGB stars are plotted in black in the upper-left and lower-right panel, respectively. The shaded colored histograms show the distributions for each of the five populations defined in the lower-left panel.}
          \label{seleRGBs}
   \end{figure*}
\vspace{-0.3 truecm}   

   \end{document}